\documentclass[aps,prl,showpacs]{revtex4}

\begin{document}

\title{Quantum gravity, minimum length and Keplerian orbits}

\author{Zurab~Silagadze}
\email{Z.K.Silagadze@inp.nsk.su}\affiliation{ Budker Institute of
Nuclear Physics and Novosibirsk State University, Novosibirsk 630
090, Russia }

\begin{abstract}
We conjecture that the modified commutation relations suggested in the 
context of quantum gravity (QG) persist also in the classical limit, if 
the momentum of the classical object is not too large, and calculate the
corresponding perihelion precession rate for Keplerian orbits. The main 
result obtained in this letter is not new. However the derivation is much 
simpler than the one proposed by Benczik et al. in Phys.\ Rev.\  D {\bf 66}, 
026003 (2002) where the corresponding precession rate was calculated for 
the first time. Our interpretation of the result is also quite different.

\end{abstract}




\maketitle

Both string theory \cite{String} and some heuristic quantum gravity 
models \cite{heuristic} imply the existence of minimum length and
the corresponding modification of the uncertainty  relations near the Planck 
scale. 

When the system has several degrees of freedom, the minimum-length
modified commutation relations have the form \cite{KMM}
\begin{equation}
\label{modcomm} 
[\hat{x}_i,\hat{p}_j]=i\hbar
\left(\delta_{ij}+\beta\hat{p}^{\,2}\delta_{ij} +
\beta^\prime\hat{p}_i\hat{p}_j \right)~.
\end{equation} 
The minimum length which follows from these commutation relations is (for 
more details see \cite{KMM,Brau})
$\delta x_\mathrm{min}=\hbar\sqrt{3\beta+\beta^\prime}~.$
 
In the particular case $\beta^\prime= 2\beta$, the realization of this 
modified Heisenberg algebra to the linear order in  $\beta$ has the form
\cite{Brau,DasVagenas}
$$ \hat{x}_i =  \hat{x}^0_i~,~~ ~~ \hat{ p}_i = \hat{ p}^0_i \left[ 1 + 
{\beta} \left(\hat{\mathbf p}^0\right)^2 \right] ~,$$
where $\hat{x}^0_i$ and $\hat{p}^0_j$ are the standard low-energy position 
and momentum operators with 
$ \left[ \hat{x}^0_i, \hat{p}^0_j \right] = i\hbar \delta_{ij}$. 
Therefore, up to this accuracy, one gets an universal QG correction to the
Hamiltonian \cite{Brau,DasVagenas}
\begin{equation} 
\hat{\mathcal{H}}  =  \frac{\hat{\mathbf p}^2}{2m} + 
V(\hat{\mathbf x}) =\frac{\left(\hat{\mathbf p}^0\right)^2}{2m} + 
V(\hat{\mathbf x}^0)
+ \frac{\beta}{m}\left(\hat{\mathbf p}^0\right)^4 + O(\beta^2)~.
\end{equation}
To consider the classical limit, the quantum mechanical commutators should be 
replaced by Poisson brackets according to
$$\frac{1}{i\hbar}[\hat A,\hat B] \to \{A,B\}.$$
Therefore we are left with the canonical conjugate variables
$$ \left \{ x_i, p_j \right \} = \delta_{ij},\;\;\; 
\left \{ x_i, x_j \right \} = 0,\;\;\; 
\left \{ p_i, p_j \right \} = 0, $$
and the perturbed Hamiltonian 
\begin{equation} 
{\mathcal{H}}  = {\mathcal{H}}_0+ \frac{\beta}{m}
p^4,
\label{pertH}
\end{equation}
where for Keplerian orbits
$${\mathcal{H}}_0=\frac{p^2}{2m}-\frac{\alpha}{r}.$$
In the above expressions we have dropped the upper index zero in the classical
limits of  $\hat{x}^0_i$ and $\hat{p}^0_j$. 

Because of the perturbation, the perihelion of the orbit begins to precess.
The precession rate is convenient to calculate by using Hamilton vector
\cite{HamiltonPP} whose precession rate coincides with the precession rate of
perihelion. The Hamilton vector has the form \cite{Hamilton}
\begin{equation}
\vec{u}=\frac{\vec{p}}{m}-\frac{\alpha}{L}\,\vec{e}_\varphi
\label{Hvector}
\end{equation}
and it is conserved in the absence of perturbation:
$$\dot{\vec{u}}=\left \{ \vec{u},{\mathcal{H}}_0\right \}=0.$$
Here $L=mr^2\dot \varphi$ is the orbital momentum and 
\begin{equation}
\vec{e}_\varphi=\frac{\vec{L}}{L}\times \frac{\vec{r}}{r}
\label{ephi}
\end{equation} 
is the unit vector in the direction of the polar angle $\varphi$ in the orbit 
plane.

When the perturbation is present, the Hamilton vector is no longer conserved:
$$\dot{\vec{u}}=\left \{ \vec{u},{\mathcal{H}}\right \}=\frac{\beta}{m}
\left\{\vec{u},p^4\right\}=\frac{2\beta}{m}p^2
\left\{\vec{u},p^2\right\}.$$
Substituting (\ref{Hvector}) and (\ref{ephi}), we get
\begin{equation}
\dot{\vec{u}}=\frac{4\beta\alpha p^2}{m L^2 r^3}\,\vec{L}\times \left (
\vec{r}\times\vec{L}\right )=\frac{4\beta\alpha p^2}{m r^3}\,\vec{r}.
\label{dotu}
\end{equation}
The precession rate of the vector $\vec{u}$ is given by \cite{Stewart}
$$\vec{\omega}=\frac{\vec{u}\times\dot{\vec{u}}}{u^2}.$$
By using $(\vec{L}\times\vec{r})\times \vec{r}=-r^2\vec{L}$ and
$u=\frac{\alpha e}{L}$, where $e$ is the eccentricity of the orbit, we get
\begin{equation}
\omega=\frac{4\beta p^2 L}{m r^3 e^2} \left (r-\frac{L^2}{m\alpha} \right).
\label{omega}
\end{equation}
Therefore, under the complete orbital cycle the Hamilton vector and hence
the perihelion of the orbit revolves by an angle 
\begin{equation}
\Delta \Theta_p=\int\limits_0^T\omega\,dt=\int\limits_0^{2\pi}
\frac{\omega}{\dot{\varphi}}\,d\varphi.
\label{Thetap}
\end{equation}
While integrating (\ref{Thetap}), to the first order in the perturbation we 
can use relations which are valid for the unperturbed orbit 
$$\frac{R}{r}=1+e\cos{\varphi}$$ and
$$\frac{p^2}{2m}-\frac{\alpha}{r}=-\frac{\alpha}{2a}.$$
Here 
$$R=\frac{L^2}{m\alpha}$$
is the semi-latus rectum of the unperturbed orbit and
$$a=\frac{R}{1-e^2}$$
is its semi-major axis.

As a result, we finally get
\begin{equation}
\Delta \Theta_p=-\frac{4\beta m\alpha}{eR}\int\limits_0^{2\pi}
\cos{\varphi}\,(1+e^2+2e\cos{\varphi})\,d\varphi=
-\frac{8\pi\beta m\alpha}{R}.
\label{ThetapF}
\end{equation}
In this formula $\alpha=G_NmM$ and $\beta=(\delta x_\mathrm{min})^2/
5\hbar^2$. Therefore, introducing the Planck mass
$$m_P=\sqrt{\frac{\hbar c}{G_N}},$$
equation (\ref{ThetapF}) can be rewritten in the form
\begin{equation}
\Delta \Theta_p=-\frac{8\pi}{5}\,\frac{\delta x_\mathrm{min}}{R}\,
\left (\frac{m}{m_P}\right )^2\,\frac{\delta x_\mathrm{min}}
{\hbar/Mc}.
\label{ThetapFF}
\end{equation}
At first sight the effect can be hugely increased by simply increasing the 
mass $m$ of the satellite body. However it should be understood that the 
r.h.s. of the modified commutation relations (\ref{modcomm}) is, in fact, 
only a truncation of the true commutation relations to lowest order terms, 
with higher order contributions neglected \cite{Berger}. This truncation to be 
valid one needs $\beta \,p^2\ll 1$, or $p\ll \hbar/\delta x_\mathrm{min}$. 
Usually it is assumed that $\delta x_\mathrm{min}$ is of the order of Planck 
length 
$$l_P=\sqrt{\frac{\hbar G_N}{c^3}}\approx 1.6\times 10^{-35}~\mathrm{m}.$$
Then $\hbar/l_P\approx 6.6~\mathrm{kg \, m/s}$. Of course, for elementary 
particles or atomic systems this is a huge momentum, but not for macroscopic 
bodies. For example, for Earth-bound satellites we need the mass of the 
satellite to be smaller than about one-tenth of a gramme our above analysis 
to make any sense. However, for the satellites light enough we expect the 
neglected terms in the true commutation relations to be indeed irrelevant and 
we get for Earth-bound satellites 
\begin{equation}
\Delta \Theta_p\approx -7 \left (\frac{6400~\mathrm{km}}{R}\right )
\left (\frac{m}{0.1~\mathrm{g}}\right )^2 10^{-2}.
\label{ThetapS}
\end{equation}

Note that similar investigations in the framework of non-commutative 
geometry were performed in \cite{NC}. Like our case, it was found that the 
modifications of physics at small scales have rather profound effect on
classical physics at large scales, something similar to the UV/IR
mixing \cite{UVIR}.

After this work was nearly completed, we learned about the paper 
\cite{Benczik} where the effects of modified commutation relations on the
classical Keplerian orbits were also considered by different method. Our
final result (\ref{ThetapF}) for perihelion precession perfectly coincides 
with the result of \cite{Benczik} (Eq. 66) when $\beta^\prime=2\beta$. 
However our derivation, being based on the Hamilton vector \cite{HamiltonPP},
is much simpler and conclusions are different. Besides, a general case 
$\beta^\prime\ne2\beta$, considered in \cite{Benczik}, assumes non-commuting 
spatial coordinates and hence non-commutative geometry. This brings some 
subtlety in the formalism and we think our independent calculation
provides a valuable cross-check of their results.

It was concluded in \cite{Benczik} that the observed precession of the 
perihelion of Mercury places a severe constraint on the value of the
minimum length which should be thirty-three (!) orders of magnitude below 
the Planck length not to have any observable consequences. We think such 
a limit looks rather strange. In our opinion the Mercury perihelion 
precession cannot be used for this goal because the truncated commutation 
relations (\ref{modcomm}) are no longer valid for such large momentum.

Nevertheless, if the involved momenta are not very large, we expect the 
modified commutation relations (\ref{modcomm}) to be valid and imply an 
observable modification of classical dynamics of macroscopic/mesoscopic 
objects (compare with $20~\mathrm{keV}=20~\mathrm{keV}$ insight from 
\cite{Strodolsky} that a cryodetector developed for dark matter searches 
can be used for mass spectroscopy with macromolecules because, being a kind 
of calorimeter, it doesn't care whether $20~\mathrm{keV}$ energy comes from
an electron or from a huge, slow, $20~\mathrm{keV}$ protein). Does this mean 
that the quantum gravity effects can be studied in space-based or even in 
table-top experiments?

In fact, it is not even excluded that such effects were already observed. 
We mean the notorious Pioneer anomaly \cite{Pioneer,Lammerzahl} and flyby 
anomalies \cite{Flyby,Lammerzahl}. However, the momentum scales involved in 
this phenomena are higher than the critical  value $6.6~\mathrm{kg\, m/s}$ 
mentioned above which does not allow us to use the classical limit of 
truncated commutation relations (\ref{modcomm}) for their analysis. 
Nevertheless, one cannot a priory exclude that the modification of Newtonian 
dynamics due to true QG commutation relations, although being much smaller 
than that follows from the incorrect use of truncated commutation relations, 
is still big enough to be relevant for these gravitational anomalies. 

Anyway, it seems worthwhile to further scrutinize the classical limit of the
KMM quantum mechanics \cite{KMM}. A naive and straightforward approach used 
in this paper, as well as in \cite{Benczik,Benczik1}, implies a deformation 
of Newtonian dynamics which can be experimentally tested either in experiments 
aimed to test Newton's second law \cite{SecondLaw,Lammerzahl1}, or in 
precision tests of the equivalence principle, because the equivalence 
principle is expected to be dynamically violated under such a deformation 
\cite{Benczik1}. 

Another place where such a deformation of Newtonian dynamics can reveal 
itself is interstellar or interplanetary dust dynamics \cite{Benczik1}
for which truncated commutation relations should be a good approximation.
Note, however, that the dust dynamics is a complicated problem involving,
besides gravity, a number of physical effects, such as  direct radiation 
pressure and stellar wind pressure, Poynting-Robertson and pseudo 
Poynting-Robertson forces, sublimation and mutual collisions of dust
grains, dust grain charging and the corresponding influence of magnetic 
fields due to Lorentz force \cite{dust}.

\section*{Acknowledgments}
The work is supported in part by grants Sci.School-905.2006.2 and 
RFBR 06-02-16192-a. The author thanks Michael Maziashvili for the useful 
comments.

\end{document}